# Dynamic Mechanism Design for Markets with Strategic Resources

**Swaprava Nath**
Indian Institute of Science
Bangalore, India
swaprava@gmail.com

**Onno Zoeter**
Xerox Research Centre Europe
Meylan, France
onno.zoeter@xrce.xerox.com

**Yadati Narahari**
Indian Institute of Science
Bangalore, India
hari@csa.iisc.ernet.in

**Christopher R. Dance**
Xerox Research Centre Europe
Meylan, France
chris.dance@xrce.xerox.com

## Abstract

The assignment of tasks to multiple resources becomes an interesting game theoretic problem, when both the task owner and the resources are strategic. In the classical, non-strategic setting, where the states of the tasks and resources are observable by the controller, this problem is that of finding an optimal policy for a Markov decision process (MDP). When the states are held by strategic agents, the problem of an efficient task allocation extends beyond that of solving an MDP and becomes that of designing a mechanism. Motivated by this fact, we propose a general mechanism which decides on an allocation rule for the tasks and resources and a payment rule to incentivize agents' participation and truthful reports.

In contrast to related dynamic strategic control problems studied in recent literature, the problem studied here has *interdependent values*: the benefit of an allocation to the task owner is not simply a function of the characteristics of the task itself and the allocation, but also of the state of the resources. We introduce a dynamic extension of Mezzetti's two phase mechanism for interdependent valuations. In this changed setting, the proposed dynamic mechanism is efficient, within period ex-post incentive compatible, and within period ex-post individually rational.

## 1 Introduction

Let us consider the example of an organization having multiple sales and production teams. The sales teams receive project contracts, which are to be executed by the production teams. The problem that the management of both teams faces is that of assigning the projects (tasks) to production teams (resources). If the efficiency levels of the production teams and the workloads of the projects are observable by the management (controller), this problem reduces to an MDP, which has been well studied in the literature (Bertsekas, 1995; Puterman, 2005). Let us call the efficiency levels and workloads together as *states* of the tasks and resources.

However, the states are usually observed privately by the individual teams (agents), who are rational and intelligent. They, therefore, might strategically misreport to the controller to increase their net returns. Hence, the problem changes from a *completely* or *partially* observable MDP into a *dynamic game* among the agents. We will consider cases where the solution of the problem involves monetary transfer. Hence, the task owners pay the resources to execute their tasks. A socially efficient mechanism would demand truthfulness and voluntary participation of the agents in this setting.

The reporting strategy of the agents and the decision problem of the controller is dynamic since the state of the system is varying with time. In addition, the above problem has two characteristics, namely, *interdependent values*: the task execution generates values to the task owners that depend on the efficiencies of the assigned resources, and *exchange economy*: a trade environment where both buyers (task owners) and sellers (resources) are present. An exchange economy is also referred to as a *market*.

The above properties have been investigated separately in literature on *dynamic mechanism design*. Bergemann and Välimäki (2010) have proposed an efficient mechanism called the *dynamic pivot mechanism*, which is a generalization of the Vickrey-Clarke-Groves (VCG) mechanism (Vickery, 1961; Clarke, 1971; Groves, 1973) in a dynamic setting, which also serves to be truthful and efficient. Athey and Segal (2007) consider a similar setting with an aim to find an efficient mechanism that is budget balanced. Cavallo et al. (2006) develop a mechanism similar to the dy-namic pivot mechanism in a setting with agents whose type evolution follows a Markov process. In a later work, Cavallo et al. (2009) consider the *periodically inaccessible* agents and dynamic private information jointly. Even though these mechanisms work for an exchange economy, they have the underlying assumption of *private values*, i.e., the reward experienced by an agent is a function of the allocation and her own private observations. Mezzetti (2004, 2007), on the other hand, explored the other facet, namely, interdependent values, but in a static setting, and proposed a truthful mechanism.

In this paper, we propose a dynamic mechanism which combines the flavors of *interdependent values* with *exchange economy* that applies to the strategic resource assignment problem. It extends the results of Mezzetti (2004) to a dynamic setting, and serves as an efficient, truthful mechanism, where agents receive non-negative payoffs by participating in it. The key feature that distinguishes our model and results from that of the existing dynamic mechanism literature is that we address the interdependent values and exchange economy together. We model dependency among agents through the dependency in value functions, while the types (private information) evolve independently. This novel model captures the motivating applications more aptly, though there are other ways of modeling dependency, e.g., through interdependent type transitions (Cavallo et al., 2009; Cavallo, 2008). We also provide simulations of our mechanism and carry out an experimental study to understand which set of properties is satisfiable in this setting. We emphasize that the assignment of tasks to strategic resources serves as a strong motivation for this work, but the scope of the mechanism applies to a more general setting as mentioned above.

The rest of the paper is organized as follows. We motivate the problem and review relevant definitions from mechanism design theory in Section 2. We introduce the model in Section 3, and present the main results in Section 4. In Section 5, we illustrate the properties of our mechanism using simulations. We conclude the paper in Section 6 with some potential future works.

## 2 Model and Background

The problem of dynamic assignment of tasks to resources can be thought of choosing a subset from a pool of tasks and resources, indexed by $N = \{0, 1, \ldots, n\}$, available at all time steps $t = 1, 2, \ldots$. The time-dependent state of each resource or characteristics (workload) of each task is denoted by $\theta_{i,t} \in \Theta_i$ for $i \in N$. We will use the shorthands $\theta_t = (\theta_{0,t}, \theta_{1,t}, \ldots, \theta_{n,t}) = (\theta_{i,t}, \theta_{-i,t})$, where $\theta_{-i,t}$ denotes the state vector of all agents excluding agent $i$, $\theta_t \in \Theta = \times_{i \in N} \Theta_i$. We will refer to $\theta_t$ as the state and to task owners and resources jointly as *agents*.

The action to be taken by the controller at $t$ is the allocation of resources to the tasks: $a_t \in A = 2^N$.

The combined state $\theta_t$ follows a first order Markov process which is governed by $F(\theta_{t+1}|a_t, \theta_t)$.

The states of the tasks and the resources $\theta_0, \ldots, \theta_n$ influence the individual valuations $v_i$ for $i \in N$, i.e. for both task owner and resources. This interdependence is expressed in the assumption that the valuation for $i$ is a function of the allocation and the *complete* state vector; $v_i : A \times \Theta \to \mathbb{R}$. This is in contrast to the classical independent valuations (also called *private values*) case where valuations are assumed to be only based on $i$'s own state; $v_i : A \times \Theta_i \to \mathbb{R}$.

The controller aims to maximize the sum of the valuations of task owners and resources, summed over an infinite horizon, geometrically discounted with factor $\delta \in (0, 1)$. The discount factor is the same for controller, task owner, and resources and is a common knowledge. We will use the terms *controller* and *central planner* interchangeably. If the controller would have perfect information about $\theta_t$'s, his objective would be to find a policy $\pi : \Theta \to A$ that guarantees an *efficient allocation* defined as follows.

**Definition 1** *(Efficient Allocation, EFF) An allocation rule $\pi$ is **efficient** if for all $t$ and for all state profiles $\theta_t$,*

$$\pi(\theta_t) \in \arg\max_{\gamma} \mathbb{E}_{\gamma, \theta_t} \left[ \sum_{s=t}^{\infty} \delta^{s-t} \sum_{i \in N} v_i(a_s, \theta_s) \right]$$

*where $\gamma = (a_t, a_{t+1}, \ldots)$ is any arbitrary sequence of allocations.*

The above problem is one of finding an optimal policy for an MDP for which polynomial time algorithms exist (See e.g. (Ye, 2005) for a general discussion).

**Strategic agents:** In the strategic setting the state $\theta_t$ is not observable by the controller. Instead $\theta_{i,t}$ is private information of participant $i$. In the context of mechanism design, $\theta_{i,t}$ is often referred to as the *type* of agent $i$ (Harsanyi, 1968). We will use the terms 'state' and 'type' interchangeably. Task owners and resources are asked to report their state at the start of each round. We will use $\hat{\theta}_{i,t}$ to refer to the reported value of $\theta_{i,t}$, and $\hat{\theta}_t$ as the reported vector of all agents. With the aim to ensure truthful reports (truthful in a sense specified more formally below), the controller will either pay to or receive from each agent a suitable amount of money at the end of each round. We restrict our attention to *quasi-linear utilities*, i.e., the agents' utilities can be written as a sum of value and payment. For the quasi-linear setting, a mechanism is completely specified by the allocation and payment rule. Each agent has a payoff for a given outcome. Hence, the aim of the central planner is to design assignment and payment rules that guarantee certain desirable properties described as follows.

Let us denote $\pi$ as the allocation rule and $\mathcal{P} = \{p_i\}_{i \in N}$ the payment rule for the entire infinite horizon. If the *reported* type profile at time $t$ is $\hat{\theta}_t$, while the *true* type profile is $\theta_t$, we denote $\pi(\hat{\theta}_t)$ as the sequence of allocations $(a_t, a_{t+1}, \ldots)$, $a_t \in A, \forall t$. Hence, a dynamic mechanism can be represented by $(\pi, \mathcal{P}) =: \mathcal{M}$. Given $\mathcal{M}$, agent $i$'s expected discounted value is given by, $\mathbb{V}_i^{\mathcal{M}}(\hat{\theta}_t, \theta_t) = \mathbb{E}_{\pi(\hat{\theta}_t), \theta_t}[\sum_{s=t}^{\infty} \delta^{s-t} v_i(a_s, \theta_s)]$, where the first argument of the value function on the left hand side denotes the reported types and the second denotes the true types. Similarly, total expected payment to agent $i$ is given by, $\mathbb{P}_i^{\mathcal{M}}(\hat{\theta}_t) = \mathbb{E}_{\pi(\hat{\theta}_t), \hat{\theta}_t}[\sum_{s=t}^{\infty} \delta^{s-t} p_i(a_s, \theta_s)]$. Hence the utility of agent $i$ in the *quasi-linear* setting is given by,

$$\mathbb{U}_i^{\mathcal{M}}(\hat{\theta}_t, \theta_t) = \mathbb{V}_i^{\mathcal{M}}(\hat{\theta}_t, \theta_t) + \mathbb{P}_i^{\mathcal{M}}(\hat{\theta}_t)$$

Where, the arguments of the function on the left hand side denote the reported and true types respectively. The notion of truthfulness is defined as follows.

**Definition 2** *(Within Period Ex-post Incentive Compatibility, EPIC) A mechanism $\mathcal{M} = (\pi, \mathcal{P})$ is Within Period Ex-post Incentive Compatible if for all agents $i \in N$, for all possible true types $\theta_t$, for all reported types $\hat{\theta}_{i,t}$ and for all $t$,*

$$\mathbb{U}_i^{\mathcal{M}}((\theta_{i,t}, \theta_{-i,t}), \theta_t) \geq \mathbb{U}_i^{\mathcal{M}}((\hat{\theta}_{i,t}, \theta_{-i,t}), \theta_t)$$

That is, truthful reporting becomes a Nash equilibrium. Another property, also referred to as *voluntary participation* is defined as follows.

**Definition 3** *(Within Period Ex-post Individual Rationality, EPIR) A mechanism $\mathcal{M} = (\pi, \mathcal{P})$ is Within Period Ex-post Individually Rational if for all agents $i \in N$, for all possible true types $\theta_t$ and for all $t$,*

$$\mathbb{U}_i^{\mathcal{M}}(\theta_t, \theta_t) \geq 0$$

That is, truthful reporting yields non-negative expected utility. The above definitions suffice for the results presented in this paper. For a more detailed description of mechanism design theory and a discussion of the desirable properties, see, e.g., (Shoham and Leyton-Brown, 2010, Chap. 10), (Narahari et al., 2009, Chap. 2).

## 3 An Exchange Economy Model with Interdependent Values

We note that the problem of dynamic task allocation to strategic resources discussed in the last section, falls in the broader setting of a market with *interdependent values*. The values of the task owners (buyers) are non-negative, while that of the resources (sellers) are non-positive (essentially they are costs), which constitute the environment of an *exchange economy*. We follow the definitions of the last section and slightly modify certain notation to address this more general setting of dynamic mechanism design.

Let $a_t(\hat{\theta}_t) \in A$ denote the allocation at time $t$, given the reported types $\hat{\theta}_t$. Though the allocation policy is given by $\pi$, we will show that the allocation decision can be reduced to per-stage decisions due to the Markov nature of the type evolution. Once allocated, the agents perfectly observe their values $V_i(a_t(\hat{\theta}_t), \theta_t, \omega), i \in N$, where $\omega$ is a random variable known as the *state of the world*. The task execution by resources can be affected by random factors, e.g., power cut, machine downtime etc., which are captured by $\omega \in \Omega$, assumed to be independent of the types of agents. Hence the values of the agents also become random variables. Let $v_i$ denote the expectation of $V_i$ taken over $\omega$ for all $i \in N$.

**Observation:** The instantaneous value can only depend on the types of the agents present at that given instant of the game. That is, if we remove a task or a resource $i$ from the system, its types would no longer affect the values of the remaining tasks or resources. This fact is crucial while calculating marginal contribution of an agent, when we consider VCG payments in the dynamic setting. We summarize the assumptions as follows.

**Assumptions:**

- The type profile, $\theta_t$, evolves as a first order Markov process. Hence, the transition from $\theta_t$ to $\theta_{t+1}$ given an allocation $a_t$ is completely captured by the joint stochastic kernel $F(\theta_{t+1}|a_t, \theta_t)$.

- For all $t$, the types $\theta_{i,t}$ are independent across agents $i \in N$. Moreover, the type transitions are also independent across agents, that is,

$$F(\theta_{t+1}|a_t, \theta_t) = \prod_{i \in N} F_i(\theta_{i,t+1}|a_t, \theta_{i,t})$$

where $F_i$'s are corresponding marginals. It is to be noted that the values of the agents are interdependent in this model as the value function depends on the complete type vector, while the types evolve independently.

- An agent deviates from truth only when her payoff *strictly* increases by misreporting.

- For each agent, the value function is bounded, and is zero if she is not allocated, i.e., $\forall\ i \in N, \theta_t \in \Theta$,
  - $|v_i(a_t, \theta_t)| < \infty,\ \forall a_t \in A$;
  - $v_i(\{a_t : i \notin a_t\}, \theta_t) = 0$.

- The distribution of $\omega$, given by $\Pi$, is known.

Let $p_{i,t}(\hat{\theta}_t)$ denote the payment to agent $i$ at time $t$ (negative payment to an agent means the agent is paying). The payment vector is denoted by $p_t(\hat{\theta}_t)$.

Once the controller decides the allocation and payment, each agent $i$ experiences her payoff, denoted by $U_i(a_t(\hat{\theta}_t), p_t(\hat{\theta}_t), \theta_t, \omega)$, and computed as the expected discounted sum of the stage-wise utilities, where the utility of each stage is the sum of value and payments. The formal expression is presented in the following section.

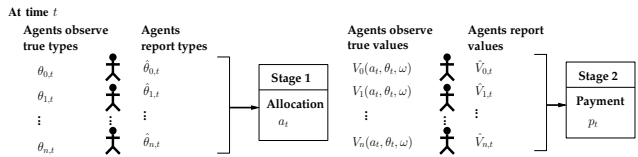

Figure 1: Graphical illustration of the proposed dynamic mechanism.

A graphical illustration of the model is given in Figure 1, with the decision of allocation and payment split into two stages. In the next section, we will explain the need for such a split.

## 4 The Generalized Dynamic Pivot Mechanism

Before presenting the proposed mechanism, let us look at other possible approaches to solve the problem by discussing a naïve attempt to decide the allocation and the payment.

**Fixed payment mechanism:** A candidate approach for the above setting is to select a set of agents and make fixed payment to them. The allocation can be done using the performance history of the agent. But one can immediately notice that this fixed payment mechanism would not be truthful. Since the payment is fixed once an agent is selected, the agents would report a type which would maximize their chance of getting selected. As the allocation is based on the history of the performance, which is common knowledge, the agents with better history would exploit the system by misreporting their type.

The above candidate mechanism suggests that, to achieve efficient allocation in a dynamic setting, one needs to consider the expected future evolution of the types of the agents, which would reflect in the allocation and payment decisions. In addition, we have interdependent values among the agents in our setting. The reason the above two approaches do not work is not an accident. Even in a static setting, if the allocation and payment are decided simultaneously interdependent valuations, one cannot guarantee efficiency and incentive compatibility together (Jehiel and Moldovanu, 2001). This compels us to split the decisions of allocation and payment in two separate stages. We would mimic the two-stage mechanism of Mezzetti (2004) for each round of the dynamic setting as follows (also see Figure 1).

### 4.1 Stage A: Type Reporting Stage
At round $t$, every agent observes her true type $\theta_{i,t}$, $i \in N$. The agents are asked to report their types. They report $\hat{\theta}_{i,t}$, $i \in N$, which may be different from their true types due to the agents' strategic nature. At the end of this stage allocation decision, $a_t \in A$ is made.

### 4.2 Stage B: Value Reporting Stage
The allocated agents now observe their values, $V_i(a_t, \theta_t, \omega)$, $i \in N$, which is a function of the true type profile and the state of the world. We assume that this observation is *perfect* for the agents. In this stage, the agents are asked to report their observed values. They report $\hat{V}_{i,t}$, $i \in N$. For this stage, we would also assume that if the agents cannot strictly gain by misreporting, they report their true observed values. At the end of this stage, the payment $p_{i,t}(\hat{\theta}_t, \hat{V}_t)$ is made to the agent $i$, $i \in N$. Immediately following the proof of the main result, we discuss why the standard dynamic pivot mechanism does not ensure IC while a two stage dynamic pivot mechanism can.

### 4.3 The Allocation and Payment Rule
Given the above dynamics of the game, the task of the central planner is to design the allocation and payment rules. We note that the expected value of agent $i$ (expected over the state of the world) is given by,

$$v_i(a_t, \theta_t) = \int_\Omega V_i(a_t, \theta_t, \omega) d\Pi(\omega) \quad (1)$$

where $V_i(a_t, \theta_t, \omega)$ is the *realized* value of the agent given the *realized* state of the world $\omega$. The objective of the social planner is to maximize social welfare given the type profile $\theta_t$, which is defined as,

$$W(\theta_t)$$
$$= \max_{\pi_t} \mathbb{E}_{\pi_t, \theta_t} \left[ \sum_{s=t}^{\infty} \delta^{s-t} \sum_{i \in N} v_i(a_s, \theta_s) \right]$$
$$= \max_{a_t} \mathbb{E}_{a_t, \theta_t} \left[ \sum_{i \in N} v_i(a_t, \theta_t) + \delta \mathbb{E}_{\theta_{t+1}|a_t, \theta_t} W(\theta_{t+1}) \right]$$

where $\pi_t = (a_t, a_{t+1}, \dots)$ is the sequence of actions starting from $t$. We denote the maximum social welfare excluding agent $i$ by $W_{-i}(\theta_t)$, which is same as the above equation except the inner sum of the first equality is over all agents $j \neq i$. Note that, due to the infinite time horizon and the stationary transition model, there exists a *stationary* policy which maximizes the social welfare, i.e., $\forall\, \theta_t \in \Theta_t$, $\exists\, a^*(\theta_t)$ such that,

$$a^*(\theta_t) \in \arg\max_{a_t} \mathbb{E}_{a_t, \theta_t}\Big[ \sum_{i \in N} v_i(a_t, \theta_t)$$
$$+ \delta \mathbb{E}_{\theta_{t+1}|a_t, \theta_t} W(\theta_{t+1}) \Big] \quad (2)$$

In the following, we propose the generalized (two-stage) dynamic pivot mechanism (GDPM), where we choose the allocation to be efficient and the payment in a manner such that it guarantees incentive compatibility and individual rationality.

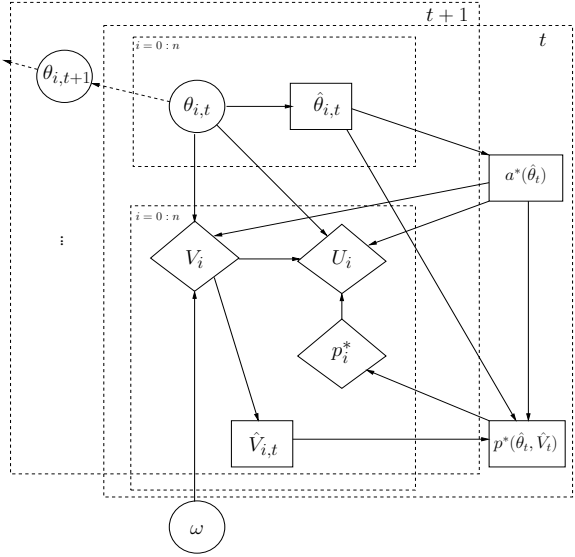

Figure 2: Multi-agent influence diagram of the proposed mechanism. Note how the two plates ranging over $i = 0 : n$ encode that the valuation of a single agent depends on the types of all agents.

**Mechanism 1 (GDPM)** *Given the reported type profile $\hat{\theta}_t$, choose the agents $a^*(\hat{\theta}_t)$ according to Equation 2. Transfer to agent $i$ after agents report $\hat{V}_t$, a payment*

$$p_i^*(\hat{\theta}_t, \hat{V}_t) = \left(\sum_{j \neq i} \hat{V}_{j,t}\right) + \delta \mathbb{E}_{\theta_{t+1}|a^*(\hat{\theta}_t), \hat{\theta}_t} W_{-i}(\theta_{t+1}) \\ - W_{-i}(\hat{\theta}_t). \quad (3)$$

The payment is similar to that of the dynamic pivot mechanism, with the difference that the first term consists of reported values. It is constructed in a way such that the expected discounted utility of an agent at true report becomes her marginal contribution. Due to space limitation, we skip the detailed discussion of the choice of such a payment, which is similar to that in Bergemann and Välimäki (2010). The mechanism can be graphically represented using a multi-agent influence diagram (MAID) (Koller and Milch, 2003), as shown in the Figure 2. We summarize the dynamics in Mechanism 1. In the following, we present the main result of the paper.

**Theorem 1** *GDPM is efficient, within period ex-post incentive compatible, and within period ex-post individually rational.*

**Mechanism 1** GDPM
> **for all** time instants $t$ **do**
>   **Stage A:**
>   **for** agents $i = 0, 1, \ldots, n$ **do**
>     agent $i$ observes $\theta_{i,t}$;
>     agent $i$ reports $\hat{\theta}_{i,t}$;
>   **end for**
>   compute allocation $a^*(\hat{\theta}_t)$ according to Eq. 2;
>   state of the world $\omega$ realizes;
>   **Stage B:**
>   **for** agents $i = 0, 1, \ldots, n$ **do**
>     agent $i$ observes $V_i(a^*(\hat{\theta}_t), \theta_t, \omega)$;
>     agent $i$ reports $\hat{V}_{i,t}$;
>   **end for**
>   compute payment to agent $i$, $p_i^*(\hat{\theta}_t, \hat{V}_t)$, Eq. 3;
>   types evolve $\theta_t \to \theta_{t+1}$ according to a first-order Markov process;
> **end for**

**Proof:** Clearly, given true reported types, the allocation is efficient by design. To show that this mechanism is indeed truthful, we need to prove that it is within period ex-post incentive compatible (EPIC). To prove EPIC, we need to consider only unilateral deviations, see Definition 2. Let us assume, all agents except agent $i$ report their true types. Hence, $\hat{\theta}_t = (\hat{\theta}_{i,t}, \theta_{-i,t})$. So, the discounted utility to agent $i$ at $t$ given the realized state of the world $\omega$ is,

$$\begin{aligned} & U_i((a^*(\hat{\theta}_t), p^*(\hat{\theta}_t, \hat{V}_t)), \theta_t, \omega) \\ & = \underbrace{V_i(a^*(\hat{\theta}_t), \theta_t, \omega) + p_i^*(\hat{\theta}_t, \hat{V}_t)}_{\text{present stage utility}} + \\ & \quad \underbrace{\delta \mathbb{E}_{\theta_{t+1}|a^*(\hat{\theta}_t), \theta_t}[W(\theta_{t+1}) - W_{-i}(\theta_{t+1})]}_{\text{discounted future marginal utility}} \\ & = V_i(a^*(\hat{\theta}_t), \theta_t, \omega) + \sum_{j \neq i} \hat{V}_{j,t} \\ & \quad + \delta \mathbb{E}_{\theta_{t+1}|a^*(\hat{\theta}_t), \hat{\theta}_t} W_{-i}(\theta_{t+1}) - W_{-i}(\hat{\theta}_t) \\ & \quad + \delta \mathbb{E}_{\theta_{t+1}|a^*(\hat{\theta}_t), \theta_t}[W(\theta_{t+1}) - W_{-i}(\theta_{t+1})] \text{ (cf. Eq. 3)} \end{aligned}$$

We notice that agent $i$'s payoff does not depend on her value report $\hat{V}_{i,t}$. Hence, agent $i$ has no incentive to misreport her observed valuation, and this applies to all agents. Therefore, by assumption (see Section 3), agents report their values truthfully, and we get,

$$\hat{V}_{i,t} = V_i(a^*(\hat{\theta}_t), \theta_t, \omega) \quad \forall \, i \in N \quad (4)$$

Hence,

$$\begin{aligned} & U_i(a^*(\hat{\theta}_t), p_i^*(\hat{\theta}_t, \hat{V}_t), \theta_t, \omega) \\ & = V_i(a^*(\hat{\theta}_t), \theta_t, \omega) + \sum_{j \neq i} V_j(a^*(\hat{\theta}_t), \theta_t, \omega) + \\ & \quad \delta \mathbb{E}_{\theta_{t+1}|a^*(\hat{\theta}_t), \hat{\theta}_t} W_{-i}(\theta_{t+1}) - W_{-i}(\hat{\theta}_t) + \\ & \quad \delta \mathbb{E}_{\theta_{t+1}|a^*(\hat{\theta}_t), \theta_t}[W(\theta_{t+1}) - W_{-i}(\theta_{t+1})] \quad (5) \end{aligned}$$

Now, we note that,

$$\mathbb{E}_{\theta_{t+1}|a^*(\hat{\theta}_t),\hat{\theta}_t} W_{-i}(\theta_{t+1}) = \mathbb{E}_{\theta_{t+1}|a^*(\hat{\theta}_t),\theta_t} W_{-i}(\theta_{t+1}) \quad (6)$$

This is because when $i$ is removed from the system (while computing $W_{-i}(\theta_{t+1})$), the values of none of the other agents will depend on the type $\theta_{i,t+1}$ (see observation in Section 3). And due to the independence of type transitions, $i$'s reported type $\hat{\theta}_{i,t}$ can only influence $\theta_{i,t+1}$. Hence, the reported value of agent $i$ at $t$, i.e., $\hat{\theta}_{i,t}$ cannot affect $W_{-i}(\theta_{t+1})$. Similar arguments show that,

$$W_{-i}(\hat{\theta}_t) = W_{-i}(\theta_t) \quad (7)$$

Hence, Equation 5 reduces to,
$$\begin{aligned}
&U_i(a^*(\hat{\theta}_t), (p_i^*(\hat{\theta}_t, \hat{V}_t)), \theta_t, \omega) \\
&= V_i(a^*(\hat{\theta}_t), \theta_t, \omega) + \sum_{j\neq i} V_j(a^*(\hat{\theta}_t), \theta_t, \omega) + \\
&\quad \cancel{\delta\mathbb{E}_{\theta_{t+1}|a^*(\hat{\theta}_t),\theta_t} W_{-i}(\theta_{t+1})} - W_{-i}(\theta_t) + \\
&\quad \delta\mathbb{E}_{\theta_{t+1}|a^*(\hat{\theta}_t),\theta_t}[W(\theta_{t+1}) - \cancel{W_{-i}(\theta_{t+1})}] \\
&\quad \text{(from Equations 6 and 7)} \\
&= \sum_{i\in N} V_i(a^*(\hat{\theta}_t), \theta_t, \omega) + \delta\mathbb{E}_{\theta_{t+1}|a^*(\hat{\theta}_t),\theta_t} W(\theta_{t+1}) \\
&\quad - W_{-i}(\theta_t) \quad (8)
\end{aligned}$$

The utility of agent $i$ given by Equation 8 depends on a specific realization of the state of the world $\omega$. It is clear that, this utility of agent $i$ is indeed random. Hence, the correct quantity to consider would be the utility expected over the state of the world, i.e.,

$$\begin{aligned}
&u_i(a^*(\hat{\theta}_t), (p_i^*(\hat{\theta}_t, \hat{V}_t)), \theta_t) \\
&= \int_\Omega U_i(a^*(\hat{\theta}_t), (p_i^*(\hat{\theta}_t, \hat{V}_t)), \theta_t, \omega) d\Pi(\omega) \quad (9)
\end{aligned}$$

Hence, the expected discounted utility to agent $i$ at $t$ is given by (using Equations 8 and 9),

$$\begin{aligned}
&u_i(a^*(\hat{\theta}_t), (p_i^*(\hat{\theta}_t, \hat{V}_t)), \theta_t) \\
&= \int_\Omega U_i(a^*(\hat{\theta}_t), (p_i^*(\hat{\theta}_t, \hat{V}_t)), \theta_t, \omega) d\Pi(\omega) \\
&= \sum_{i\in N} \int_\Omega V_i(a^*(\hat{\theta}_t), \theta_t, \omega) d\Pi(\omega) + \\
&\quad \delta\mathbb{E}_{\theta_{t+1}|a^*(\hat{\theta}_t),\theta_t} W(\theta_{t+1}) - W_{-i}(\theta_t) \\
&= \sum_{i\in N} v_i(a^*(\hat{\theta}_t), \theta_t) + \delta\mathbb{E}_{\theta_{t+1}|a^*(\hat{\theta}_t),\theta_t} W(\theta_{t+1}) \\
&\quad - W_{-i}(\theta_t), \quad \text{(from Eq. 1)} \\
&\leq \sum_{i\in N} v_i(a^*(\theta_t), \theta_t) + \delta\mathbb{E}_{\theta_{t+1}|a^*(\theta_t),\theta_t} W(\theta_{t+1}) \\
&\quad - W_{-i}(\theta_t), \quad \text{(by definition of } a^*(\theta_t), \text{Eq. 2)} \\
&= u_i(a^*(\theta_t), (p_i^*(\theta_t, V_t)), \theta_t), \\
&\quad \text{where } V_t = (V_i(a^*(\theta_t), \theta_t, \omega))_{i\in N}
\end{aligned}$$

This proves within period ex-post IC. We notice that in this equilibrium, i.e., the ex-post Nash equilibrium, the utility of agent $i$ is given by,

$$\begin{aligned}
&u_i(a^*(\theta_t), (p_i^*(\theta_t, V_t)), \theta_t) \\
&= \sum_{i\in N} v_i(a^*(\theta_t), \theta_t) + \delta\mathbb{E}_{\theta_{t+1}|a^*(\theta_t),\theta_t} W(\theta_{t+1}) \\
&\quad - W_{-i}(\theta_t) \\
&= W(\theta_t) - W_{-i}(\theta_t) \\
&\geq 0
\end{aligned}$$

This proves within period ex-post IR. ∎

**Discussion:** It is interesting to note that, if we tried to use the dynamic pivot mechanism (DPM), (Bergemann and Välimäki, 2010), unmodified in this setting, the true type profile $\theta_t$ in the summation of Eq. 5 would have been replaced by $\hat{\theta}_t$, since this comes from the payment term (Eq. 3). The proof for the DPM relies on the *private value* assumption (see para 3 of Section 2) such that, when reasoning about the valuations for the other agents $j \neq i$, we have $V_j(a^*(\hat{\theta}_t), \hat{\theta}_{i,t}, \theta_{-i,t}, \omega) = V_j(a^*(\hat{\theta}_t), \theta_{j,t}, \omega)$. But in our dependent value setting, we cannot replace $\hat{\theta}_t$ in $V_j(a^*(\hat{\theta}_t), \hat{\theta}_t, \omega)$ by $\theta_t$ for $j \neq i$, and hence the proof of IC in DPM does not work. We have to invoke the second stage of value reporting and also use Eq. 4.

**Complexity:** The non-strategic version of the resource to task assignment problem was that of solving an MDP, whose complexity was polynomial in the size of state-space (Ye, 2005). Interestingly, for the proposed mechanism, the allocation and payment decisions are also solutions of MDPs (Equations 2, 3), and we need to solve $|N|+1$ of them. Hence the proposed GDPM has polynomial time complexity in the number of agents and state-space, which is the same as that of DPM.

## 5 Simulation Results

In this section, we demonstrate the properties satisfied by GDPM through simple illustrative experiments, and compare the results with a naïve fixed payment mechanism (CONST). In addition to the already proven properties of GDPM, we also explore two more properties here. We call a mechanism *payment consistenct* (PC) if the buyer pays and seller receive payment in each round, and *budget balanced* (BB) if the sum of the monetary transfers to all the agents is non-positive (no deficit). For brevity, we choose a relatively small example, but analyze it in detail.

**Experimental Setup:** Let us analyze the example of Section 1, where we consider three agents: a project coordinator (task owner/buyer) and two production teams (sellers). Let us assume that the difficulty of the task, and the efficiency of each team can take three possible values: high (H), medium (M), and low (L), which are private information (types) of the agents. To define value functions, we associate a real number to each of these types, given by 1 (high), 0.75 (medium), and 0.5 (low). Let us call the task owner agent 0 and the two production teams agents 1 and 2. Denote the types of the agents at time $t$ by $\theta_{0,t}$, $\theta_{1,t}$, and $\theta_{2,t}$ respectively. We consider the following value structure (the $v_i$'s, expected over the state of the world)

$$v_0(a_t, \theta_t) = \left( \frac{k_1}{\theta_{0,t}} \sum_{i \in a_t, i \neq 0} \theta_{i,t} - k_2 \right) \mathbf{1}_{0 \in a_t};$$

$$v_j(a_t, \theta_t) = -k_3 \theta_{j,t}^2 \mathbf{1}_{j \in a_t}, j = 1, 2; k_i > 0, i = 1, 2, 3.$$

The value of the center is directly proportional to the sum total efficiency of the selected employees and inversely proportional to the difficulty of the task. For each production team, the value is negative (representing cost). It is proportional to the square of the efficiency level, representing a law of diminishing returns. Though the results in this paper do not place any restriction on the value structure, we have chosen a form that is consistent with the example. Note that the value of the center depends on the types of all the agents, giving rise to interdependent values. Also, because of the presence of both buyers and sellers, the market setting here is that of an exchange economy.

Type transitions are independent and follow a first order Markov chain. We choose a transition probability matrix that reflects that efficiency is likely to be reduced after a high workload round, improved after a low workload round (e.g. when a production team is not assigned).

**A Naïve Mechanism (CONST):** We consider another mechanism, where the allocation decision is the same as that of GDPM, that is, given by Equation 2 but the payment is a fixed constant $p$ if the team is selected, and the task owner is charged an amount $p$ times the number of teams selected. This mechanism satisfies, by construction, PC and BB properties. We call this mechanism CONST.

The experimental results for an infinite horizon with discount factor $\delta = 0.7$ are summarized in Figures 3, 4, and 5. There are 3 agents each with 3 possible types: the $3^3 = 27$ possible type profiles are represented along the $x$-axis of the plots. The ordering is as represented in the bottom plot of Figure 3. This plot also shows the (stationary) allocation rule assuming truthful reports: a ○ denotes the respective agent would be selected, a × it would not.

The top plot in Figure 3 shows the utility (defined in Eq. 9) to the task owner. The middle plot the utility to production team 1 (note that the production teams are symmetric, so we need to study only one). Since we are interested in ex-post equilibriums, we show utilities in the setting where all other agents report truthfully, and consider the impact of misreporting by the agent under study. Circles represent true reports, plus signs denote the utilities from a misreport. We see in Figure 3 that indeed, as it should by the ex-post IR and IC results from Section 4, utilities for truthful reports lie at or above the utility for misreports and are positive. Figure 4, which shows the analogue plots of Figure 3 for CONST, show that the naïve method is not ex-post IC (for both task owner and production teams there are misreports that yield higher utility than a truthful report).

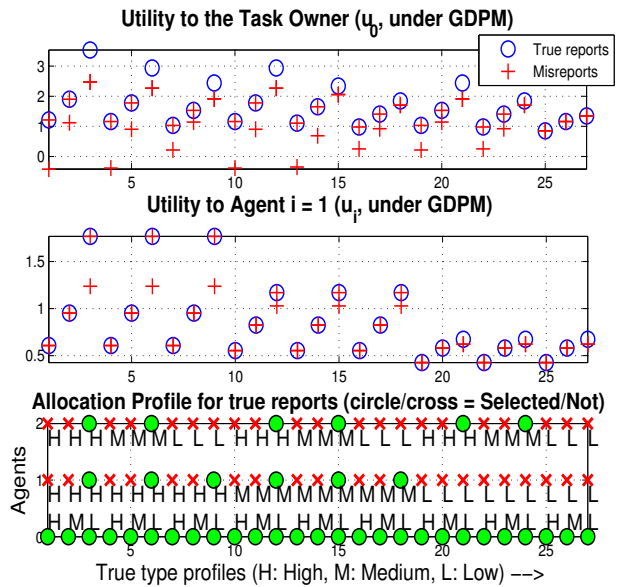

Figure 3: Utility of task owner and production team 1 under GDPM as function of true type profiles. The ordering of the $3^3 = 27$ type profiles is represented in the bottom plot.

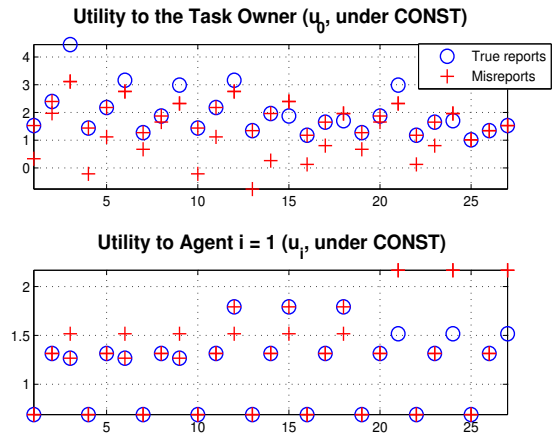

Figure 4: Utility of task owner and team 1 under CONST as function of true type profiles (same order as in Fig 3.

Figure 5 investigates the other two properties, PC and BB, for GDPM. We observe that neither are satisfied for GDPM. We summarize the results in Table 5. Not surprisingly, GDPM satisfies three very desirable properties. However, there are a few cases where it ceases to satisfy PC and BB. On the other hand, CONST, which satisfies the last two properties by construction, fails to satisfy the others. It seems that satisfying all of these properties together may be impossible in a general dependent valued exchange economy, a result similar to the Myerson-Satterthwaite theorem. However, it is promising to derive bounds on payment inconsistency and budget deficit for a truthful mechanism such as GDPM, which we leave as potential future work.

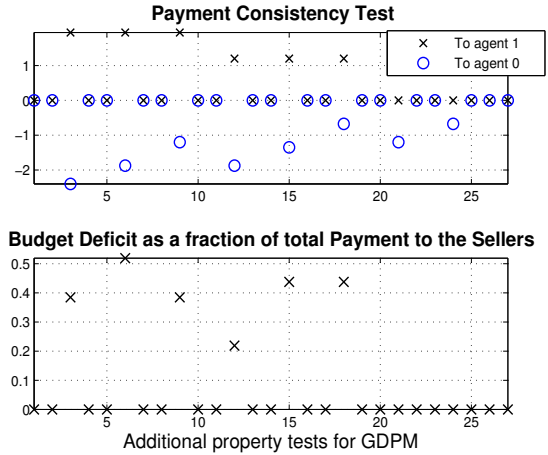

Figure 5: Payment consistency and budget properties of GDPM. The x-axis follows same true profile order as in Fig 3

|       | EFF | EPIC | EPIR | PC | BB |
|-------|-----|------|------|----|----|
| GDPM  | ✓   | ✓    | ✓    | ×  | ×  |
| CONST | ×   | ×    | ×    | ✓  | ✓  |

Table 1: Simulation summary

## 6  Conclusions and Future Work

In this paper, we have formulated the dynamic resource to task allocation problem as a dynamic mechanism design problem and analyzed its characteristics. Our contributions are as follows.

- We have proposed a dynamic mechanism, that is efficient, truthful and individually rational in a *dependent-valued exchange economy*.
- The mechanism comes at the similar computational cost as that of its non-strategic counterpart.
- We have illustrated and explored the properties exhibited by our mechanism through simulations, and observed that the proposed mechanism does not satisfy EFF, EPIC, EPIR, PC, and BB at the same time. A precise characterization of the (im)possibilities for the model we studied would be desirable.

Other directions for future work would be to design mechanisms having additional desirable properties in this setting. If the impossibility result is true, then the bounds on the payment inconsistency and budget imbalance need to be explored. We can weaken the notion of truthfulness to get a better handle on the other properties. In contrast to the interdependent values, the case of dependent types and type transitions are also interesting models of further study.

## 7  Acknowledgments

We would like to thank David C. Parkes for useful discussions, and Pankaj Dayama for discussions and comments on the paper.

## References

Susan Athey and Ilya Segal. An efficient dynamic mechanism. Working Paper, Stanford University, Earlier version circulated in 2004, 2007.

Dirk Bergemann and Juuso Välimäki. The Dynamic Pivot Mechanism. *Econometrica*, 78:771–789, 2010.

Dimitri P. Bertsekas. *Dynamic Programming and Optimal Control*, volume I. Athena Scientific, 1995.

Ruggiero Cavallo. *Social Welfare Maximization in Dynamic Strategic Decision Problems*. PhD thesis, School of Engineering and Applied Sciences, Harvard University, Cambridge, Massachusetts, May 2008.

Ruggiero Cavallo, David C. Parkes, and Satinder Singh. Optimal coordinated planning amongst self-interested agents with private state. In *Proceedings of the Twenty-second Annual Conference on Uncertainty in Artificial Intelligence (UAI 2006)*, pages 55–62, 2006.

Ruggiero Cavallo, David C. Parkes, and Satinder Singh. Efficient Mechanisms with Dynamic Populations and Dynamic Types. Technical report, Harvard University, 2009.

Edward Clarke. Multipart Pricing of Public Goods. *Public Choice*, (8):19–33, 1971.

Theodore Groves. Incentives in Teams. *Econometrica*, 41(4):617–31, July 1973.

John C. Harsanyi. Games with Incomplete Information Played by "Bayesian" Players, I-III. Part II. Bayesian Equilibrium Points. *Management Science*, 14(5):pp. 320–334, 1968.

P. Jehiel and B. Moldovanu. Efficient Design with Interdependent Valuations. *Econometrica*, (69):1237–1259, 2001.

Daphne Koller and Brian Milch. Multi-agent influence diagrams for representing and solving games. *Games and Economic Behavior*, Volume 45(Issue 1):181–221, October 2003.

Claudio Mezzetti. Mechanism Design with Interdependent Valuations: Efficiency. *Econometrica*, 2004.

Claudio Mezzetti. Mechanism design with interdependent valuations: surplus extraction. *Economic Theory*, 31:473–488, 2007.

Y. Narahari, Dinesh Garg, Ramasuri Narayanam, and Hastagiri Prakash. *Game Theoretic Problems in Network Economics and Mechanism Design Solutions*. Springer-Verlag, London, 2009.

Martin L. Puterman. *Markov Decision Processes: Discrete Stochastic Dynamic Programming*. Wiley Interscience, 2005.

Yoav Shoham and Kevin Leyton-Brown. *Multiagent Systems: Algorithmic, Game-Theoretic, and Logical Foundations*. Cambridge University Press, 2010.

William Vickery. Counterspeculation, Auctions, and Competitive Sealed Tenders. *Journal of Finance*, 1961.

Yinyu Ye. A New Complexity Result on Solving the Markov Decision Problem. *Math. Oper. Res.*, 30:733–749, August 2005.